\documentclass[10pt,conference]{IEEEtran}
\usepackage{epsfig,rotating,setspace,latexsym,amsmath,epsf,amssymb,amsfonts,bm,theorem,cite,enumerate,longtable,accents,float,physics}
\usepackage{algorithm,algorithmic,graphicx,epsf,authblk,epstopdf,url,xcolor, soul,multirow,bbm,relsize}
\usepackage{mathtools,comment}
\usepackage[center]{qtree}
\usepackage{tree-dvips}
\usepackage[linguistics]{forest }

\newtheorem{theorem}{Theorem}
\newtheorem{corollary}{Corollary}
\newtheorem{definition}{Definition}
\newtheorem{remark}{Remark}
\newtheorem{lemma}{Lemma}

\newenvironment{Proof}[1]{\medskip\par\noindent{\bf Proof:\,}\,#1}{{\mbox{\,$\blacksquare$}\par}}

\IEEEoverridecommandlockouts

\DeclareMathOperator{\diag}{diag}

\allowdisplaybreaks

\title{Quantum $X$-Secure $B$-Byzantine $T$-Colluding Private Information Retrieval}
\author{Mohamed Nomeir \qquad Alptug Aytekin \qquad Sennur Ulukus\\
	\normalsize Department of Electrical and Computer Engineering\\
	\normalsize University of Maryland, College Park, MD 20742 \\
	\normalsize  \emph{mnomeir@umd.edu} \qquad \emph{aaytekin@umd.edu} \qquad \emph{ulukus@umd.edu}}
 
\begin{document}

\maketitle

\begin{abstract}
   We consider the problems arising from the presence of Byzantine servers in a quantum private information retrieval (QPIR) setting. This is the first work to precisely define what the capabilities of Byzantine servers could be in a QPIR context. We show that quantum Byzantine servers have more capabilities than their classical counterparts due to the possibilities created by quantum encoding procedures. We focus on quantum Byzantine servers that can apply any reversible operation on their individual qudits. In this case, Byzantine servers can generate any error, i.e., this covers \emph{all} possible single qudit operations that can be applied by Byzantine servers on their qudits. We design a scheme based on cross-subspace alignment (CSA) and we show that this scheme achieves superdense coding gain in some cases.
\end{abstract}

\section{Introduction}
Private information retrieval (PIR) problem is introduced in the seminal paper \cite{chor}. In PIR, a user wishes to retrieve a message out of $K$ possible messages from $N$ databases without letting any database know the index of the retrieved message. The capacity, i.e., the ratio between the number of bits of the required message to the total number of downloaded bits, for this problem, for $N$ databases having $K$ replicated messages is derived in \cite{c_pir} to be $C = (1+\frac{1}{N} + \frac{1}{N^2}+\ldots+\frac{1}{N^{K-1}})^{-1}$. A variation, where any $T$ out of $N$ databases could be colluding, called $T$-colluding PIR, or TPIR in short, is studied in \cite{colluding}, and the capacity is found to be $(1+\frac{T}{N}+\ldots+(\frac{T}{N})^{K-1})^{-1}$. Another variation, where the messages need to be securely stored against any $X$ communicating databases, called XPIR, is studied in \cite{first_xsecure}. The largest achievable rate found to date for the $X$-secure TPIR, i.e., XTPIR, is $1 - \frac{X+T}{N}$, given in \cite{csa}. In \cite{byzantine_tpir}, the Byzantine variation of TPIR (BTPIR), where any $B$ of the databases may act as a Byzantine entity and manipulate the answers transmitted to the user, is studied and the capacity is found to be $\frac{N-2B}{N}(1+\frac{T}{N-2B}+\ldots+(\frac{T}{N-2B})^{K-1})^{-1}$. In \cite{jafar_byzantine}, the highest achievable rate for XBTPIR is found to be $1-\frac{X+T+2B}{N}$. A non-comprehensive list of some of the other variations of the PIR problem is \cite{arbitrarycollusion, banawan_eaves, banawan_multimessage_pir, banawan_pir_mdscoded, batuhan_hetero,C_SETPIR, ChaoTian, codedstorage_adversary_tpir, grpahbased_pir, Kumar_PIRarbCoded, multimessage_pir_sideinfo, mat_pir_byzantine_colluding, mdstpir, multiround_pir, nomeirasymmetric}; see also \cite{UlukusPIRLC}. 

The quantum version of PIR (QPIR) is introduced in \cite{qpir}. In QPIR, there are $N$ databases that share entanglement state, and store $K$ replicated classical messages, and a user, who wishes to retrieve a message privately, by sending queries to the databases over classical channels. Upon receiving the queries, the databases encode their qudits with their answers and send them over separate quantum channels. It is shown in \cite{qpir} that the capacity is double the classical capacity of symmetric PIR (SPIR) with just 2 servers, i.e., capacity equals $1$. In \cite{qtpir}, the same model is adopted for $T$-colluding databases, and it is shown that the capacity in the quantum setting is double the capacity of its classical counterpart as well. More recently, a mathematical abstraction is developed for the setting with $N$ transmitters sharing entanglement and transmitting over separate quantum channels to a common receiver, called the $N$-sum box abstraction \cite{nsumbox}. It is shown in \cite{nsumbox} that for QXTPIR we can achieve superdense coding gain, i.e., double the rate compared to their classical counterpart. Other variations can be found in \cite{qpir_colluding_mdscoded,qpir_star_product_codes,qtpir_t=n-1, our_journal}. 

In this paper, we study the Byzantine version of QXTPIR, i.e., QXBTPIR. In this setting, there are $N$ entangled servers, out of which, $B$ are Byzantine. We use the generalized Pauli $\mathsf X$ and $\mathsf Z$ for encoding and their corresponding projective value measurements (PVMs) for decoding at the user side. It is clear that, in the quantum setting, Byzantine servers can manipulate both the required transmitted dits and the encoding structure available at their side. Also, in contrast to classical Byzantine servers, quantum Byzantine servers share an entangled state with the honest servers, which implies the existence of an intertwined connection between all of them. In addition, there is a quantum encoding scheme, i.e., unitary gates, that is applied on qudits based on the symbols to be transmitted. 

In one extreme example, if the Byzantine servers apply a measurement on their state, then the entanglement state will be broken, and the maximum achievable rate will become the classical rate. Other cases include the usage of different encoding unitary operators and sending pure noise symbols, which is the matter of investigation in this paper. We remark that even if encoding is done on a single qudit by a single unitary gate, the number of possible unitary gates that can be used by the Byzantine servers is uncountably infinite. In this paper, we focus on Byzantine servers that can manipulate classical dits and can apply arbitrary Pauli operations. We note that the linear combination of Pauli gates covers \emph{arbitrary} single qudit quantum operations \cite{nielsen-chuang}. To that end, we design a scheme that combats this kind of Byzantine servers and show that the superdense coding gain is maintained by our scheme, when the number of interference terms is greater than half the number of servers, i.e., $R_Q =  2(1-\frac{X+T+2B}{N})$.

\section{Problem Formulation} \label{problem_formulation}
Before delving into the QXBTPIR formulation, some important definitions related to quantum physics and quantum information theory need to be recalled. In addition to definitions given in \cite{our_journal}, the following definitions are needed.

\begin{definition}[Quantum operation]
    A quantum operation $\mathcal{E}$ is a linear, completely-positive, and trace-preserving map from the set of all density matrices to itself.
\end{definition} 

\begin{definition}[Kraus representation] \label{kraus}
    Any quantum operation $\mathcal{E}$ acting on a quantum state $\rho$ can be written in the form,
    \begin{align}
        \mathcal{E}(\rho) = \sum_i M_i\rho M_i^{\dagger},
    \end{align}
    where $\sum_i M_i^\dagger M_i= I$.
\end{definition}

Now, consider a system consisting of $N$ databases and $K$ messages. Out of the $N$ databases, $X$ of them can communicate, i.e., share their whole dataset with each other to decode the messages, $T$ of them can collude, i.e., share the user's queries, and $B$ of them can manipulate the classical dits or manipulate the encoders available at their sides. For this quantum $X$-secure, $B$-Byzantine, $T$-colluding PIR (QXBTPIR) problem, we follow the system models introduced in the literature \cite{qtpir,qtpir_t=n-1,our_quantum_first}. The databases store $S_{n}$, $n \in [N]$, as classical dits and share an entangled state of $N$ quantum dits denoted by $\rho$. The user sends the queries $Q_{[N]}^{[\theta]}$ over a classical channel to each of the $N$ databases, and each database $n$, $n\in[N]$, with the quantum system $\mathcal{A}_n^0 = tr_{j=[1:N]\atop j \neq n}(\rho)$, where $tr(\cdot)$ is the trace operator, replies to the user queries over a separate quantum channel. Upon receiving the query, each honest database $n$ performs a quantum encoding operation based on the received query $Q_n^{[\theta]}$, storage $S_n$ and $\mathcal{A}_n^0$ to produce the quantum state $\mathcal{A}_n^{[\theta]}$ as, 
\begin{align}
    \mathcal{A}_n^{[\theta]} = Enc_n(Q^{[\theta]}_n,S_n, \mathcal{A}_n^0), \quad n \in [N]\setminus \mathcal{B}, 
\end{align}
where $Enc_n$ is the $n$th database's encoder, and $|\mathcal{B}| \leq B$. The received state at the user is given as, 
\begin{align}
    \mathcal{A}_{[N]}^{[\theta]} = \mathcal{A}_1^{[\theta]} \otimes \ldots \otimes \mathcal{A}_N^{[\theta]},
\end{align}
where $\otimes$ is the tensor product. The $K$ messages, $W_1,\ldots,W_K$, are sub-packetized into equal length $L$, that are independent and identically distributed. The variables $N$, $T$, $X$, $B$, and $K$ are assumed to be known by the user. The messages are generated uniformly at random from the field $\mathbb{F}_q$, with $q=p^r$, where $p$ is any prime number. Thus, in $q$-ary bits (dits),
\begin{align} 
    H(W_k)&=L, \quad k \in [K], \label{message_entropy}\\
    H(W_{[K]})&= \sum_{k=1}^{K} H(W_k) = KL. \label{messages_entropy}
\end{align}
The messages $W_{[K]}$ need to be secure against any $X$ communicating databases, i.e.,
\begin{align} \label{x_storage_constraint}
    I(W_{[K]}; S_{\mathcal{X}})= 0,
\end{align}
where $S_{\mathcal{X}}$ is all stored data in any set of $\mathcal{X}$ databases satisfying $|\mathcal{X}|\leq X$.

The user wants to retrieve a message $W_{\theta}$, where $\theta$ is chosen uniformly at random from $[K]$. Thus, a query $Q_n^{[\theta]}$ is sent to the $n$th database. The set of queries sent to all $N$ databases is denoted by $Q_{[N]}^{[\theta]}$. As the user is unaware of the messages, the queries generated are independent of the message content,  
\begin{align}\label{colluding_constraint}
    I(W_{[K]}; Q_{[N]}^{[\theta]}) = 0, \quad \theta \in [K].
\end{align}
In addition, we require that the index of the retrieved message by the user is secure against any $T$ colluding databases, i.e.,
\begin{align}\label{privacy_constraint}
    I(\theta; Q_{\mathcal{T}}^{[\theta]}) = 0, \quad \theta \in [K],
\end{align}
where $\mathcal{T} \subset [N], ~ |\mathcal{T}| \leq T$.

Additionally, the Von Neumann entropy $S(\cdot)$ of the required message $W_{\theta}$ given the queries and the answers must be zero for decodability (correctness),
\begin{align}\label{quantum_decodability}
    S(W_{\theta}| \mathcal{A}_{[N]}^{[\theta]},Q_{[N]}^{[\theta]})=0, \quad \theta \in [K],
\end{align}
where $W_{\theta}$ can be decoded given that there is a set $\mathcal{B}$ of Byzantine servers such that $\mathcal{B} \subset [N]$ and $|\mathcal{B}| \leq B$.

The QXBTPIR rate $R_Q$ for the retrieval scheme satisfying the above equations is defined as,
\begin{align}
    R_Q = \frac{H(W_{\theta})}{\log dim(\mathcal{A}_1^{[\theta]}\otimes\ldots\otimes\mathcal{A}_N^{[\theta]})},
\end{align}
where $dim(A)$ is the vector space dimension spanned by $A$.

In this paper, we follow the encoding and decoding structure using the $N$-sum box abstraction introduced recently in \cite{nsumbox}. In the encoding stage, the databases use generalized Pauli operators $\mathsf{X}(a) = \sum_{j=0}^{q-1} \ket{j+a}\bra{j}$, and  $\mathsf{Z}(a) = \sum_{j=0}^{q-1} \omega^{tr(aj)} \ket{j}\bra{j}$, where $q=p^r$ with $p$ a prime number, $a \in \mathbb{F}_q$ and $\omega = \exp(2\pi i /p)$. In the decoding stage, the user applies PVM defined on the quotient space of the stabilizer group $\mathcal{L}(\mathcal{V})$ defined by,
\begin{align}
    \mathcal{L}(\mathcal{V}) = \{c_{v} \Tilde{W}(v) : v \in \mathcal{V} \},
\end{align}
where $\mathcal{V}$ is a self-orthogonal subspace in $\mathbb{F}_q ^{2N}$, 
\begin{align}
     \Tilde{W}(v) = \mathsf{X}(v_1) \mathsf{Z}(v_{N+1}) \otimes \ldots \otimes \mathsf{X}(v_N) \mathsf{Z}(v_{2N}),
\end{align}
and $c_{v} \in \mathbb{C}$ is chosen such that $\mathcal{L}(\mathcal{V})$ is an Abelian subgroup of $HW_{q}^N$  with $c_{v}I_{q^N}$ being an element of the stabilizer group if $c_{v}=1$, where $HW_{q}^N$ is the Heisenberg-Weyl group defined as follows,
\begin{align}
    HW_{q}^N = \{ c \Tilde{W}(s) : s \in \mathbb{F}_q^{2N}, c \in \mathbb{C} \setminus \{0\} \}.
\end{align}
\section{Main Results} \label{main_results}
\begin{theorem} \label{main_thm}
    For the quantum $X$-secure, $B$-Byzantine, $T$-colluding  PIR (QXBTPIR), where the $B$ Byzantine servers can manipulate encoders with any single qudit quantum operation, with $N$ databases, which are allowed to share entanglement and have quantum channels for answer strings, with $K$ messages, the following rates are achievable,    
    \begin{align}
        R_Q= 2\left(1-\frac{X+T+2B}{N}\right)  
    \end{align}

    when $X+T \geq N/2$, 
    \begin{align}
        R_Q= \max\left\{\frac{N-4B}{N},1-\frac{X+T+2B}{N}\right\}
    \end{align}
    when $X+T < \frac{N}{2} \leq  N-2B$, and 
    \begin{align}
        R_Q= 1-\frac{X+T+2B}{N}
    \end{align}
    when $X+T <  N-2B < \frac{N}{2} $.
\end{theorem}

\begin{corollary} \label{cor_eaves}
For the quantum $X$-secure, $B$-Byzantine, $T$-colluding PIR, where the $B$ Byzantine servers can manipulate encoders with any single qudit quantum operation, with $N$ databases, which are allowed to share entanglement and have quantum channels for answer strings, in the presence of $E$ eavesdroppers (QXBTEPIR), with $X+M \geq \frac{N}{2}$, $X+M+2B \leq N$ and $N+E \leq 2X+2M+4B$ with $K$ messages, the following rate is achievable,
\begin{align}
    R_Q= \min\left\{1,2\left(1-\frac{X+M+2B}{N}\right)\right\},
\end{align}
where $M=\max\{E,T\}$.
\end{corollary}

\begin{remark}
    As shown in Corollary~\ref{cor_eaves}, the scheme presented here integrates with the scheme presented in \cite{our_journal} to obtain an achievable rate for the more general problem of QXBTEPIR.
\end{remark}

\section{Achievable Scheme} \label{achievable_scheme}
First, let us consider that the system consists of $N$ databases, any $X$ of them can communicate, $T$ can collude, and $B$ can be Byzantine with the two Pauli operations $\mathsf{X}$ and $\mathsf{Z}$, and $K$ messages in the whole system. That is, in this case, the Byzantine servers use the $N$-sum box as a black box and are acting Byzantine in their classical inputs to the encoders $\mathsf{X}$ and $\mathsf{Z}$, i.e., the inputs are uniform random variables in $\mathbb{F}_q$. The sub-packetization is given by $L=N-X-T-2B$. 

An important concept that needs to be explained before delving into the scheme is that any quantum operation on a single qudit, $\mathcal{E}$, can be written as a linear combination of the generalized Pauli operations, i.e., $M = \sum_{i,j \in \mathbb{F}_q} a_{ij} \mathsf X(i) \mathsf Z(j)$, noted in \cite{nielsen-chuang} and proven here for completeness. Using that concept, we can divide the problem into separate cases and apply superposition afterward. As a starting point, let us consider a retrieval of $L$ symbols in the classical case. The storage for the $n$th database is given by,
\begin{align}\label{classical_storage}
    S_n=\begin{bmatrix}
        W_{\cdot,1} + \sum_{i=1}^X(f_1-\alpha_n)^i R_{1i}\\
        W_{\cdot,2} + \sum_{i=1}^X(f_2-\alpha_n)^iR_{2i}\\
        \vdots\\
        W_{\cdot,L} + \sum_{i=1}^X(f_L-\alpha_n)^iR_{Li}\\
    \end{bmatrix},
\end{align}
where $W_{\cdot,j}=[W_{1,j},\ldots,W_{K,j}]^t$ is a vector representing the $j$th bit of all $K$ messages, with $W_{i,j}$ being the $j$th bit of message $i$, $R_{ji}$ are uniform independent random vectors with the same dimensions as $W_{\cdot,j}, ~ j \in [L]$, and $\alpha_i, f_j \in \mathbb{F}_q$ are distinct where $i\in [N]$ and $j \in[L]$, and $t$ denotes the transpose operator.

To retrieve the $L$ symbols of $W_{\theta}$, the user sends the following queries to the $n$th database,
\begin{align}\label{classical_query}
    Q_n^{[\theta]} =\begin{bmatrix}
        \frac{1}{f_1-\alpha_n}\left(e_{\theta}+\sum_{i=1}^T(f_1-\alpha_n)^i Z_{1i}\right)\\
        \vdots\\
        \frac{1}{f_L-\alpha_n}\left(e_{\theta}+\sum_{i=1}^T(f_L-\alpha_n)^i Z_{Li}\right)
    \end{bmatrix},
\end{align}
where $e_{\theta}$ is a vector of length $K$ with $1$ in the $\theta$th index and zero otherwise, and $Z_{ij}$ are uniform independent random vectors of length $K$ each, chosen by the user. Afterward, the databases perform the following operation, given the queries,
\begin{align}
    A_n^{[\theta]}= S_n^t Q_n^{[\theta]}.
\end{align}
Thus, the answers from the $N$ databases can be written as,
\begin{align}
    &A^{[\theta]} = [A_1^{[\theta]},\ldots,A_N^{[\theta]}]^t  \nonumber \\ &=\mathtt{CSA}_{N\times L+X+T}(f_{[L]},\alpha_{[N]}) [W_{\theta 1},\ldots,W_{\theta L}, I'_1,\ldots,I'_{X+T}]^t \nonumber\\ &\quad + \Delta_B, 
\end{align}
where $I'_1,\ldots,I'_{X+T}$ are interference terms, 
\begin{align}
    &\!\!\!\!\mathtt{CSA}_{N\times L+X+T}(f_{[L]},\alpha_{[N]}) \nonumber\\
    &=\begin{bmatrix}
        \frac{1}{f_1-\alpha_1}&\ldots&\frac{1}{f_L-\alpha_1}&1& \alpha_1&\ldots&\alpha_1^{X+T-1}\\
        \frac{1}{f_1-\alpha_2}&\ldots&\frac{1}{f_L-\alpha_2}&1& \alpha_2&\ldots&\alpha_2^{X+T-1}\\       \vdots&\ldots&\vdots&\vdots&\vdots&\ldots&\vdots \\
           \frac{1}{f_1-\alpha_N}&\ldots&\frac{1}{f_L-\alpha_N}&1& \alpha_N&\ldots&\alpha_N^{X+T-1}
    \end{bmatrix},
\end{align}
and $\Delta_B$ is all zero $N \times 1$ vector except at $B$ locations with random noise variables. Any $X+T+L$ rows of $\mathtt{CSA}_{N\times N-2B}(f_{[L]},\alpha_{[N]})$ is a square invertible Cauchy-Vandermonde matrix as shown in \cite{jafar_byzantine}. 

The main issue in that scheme is that the $N$-sum box abstraction cannot be applied directly \cite{nsumbox}, i.e., how the user performs measurements. To overcome that issue, consider the following simple equivalence relation
\begin{align}
    &\mathtt{CSA}_{N\times N-2B}(f_{[L]},\alpha_{[N]}) [x_1,\ldots,x_{N-2B}] + \Delta_B \nonumber\\
     &=\mathtt{CSA}_{N\times N}(f_{[L]},\alpha_{[N]}) ([x_1,\ldots,x_{N-2B}, 0_{2B}]+\Tilde{\Delta}_B), 
\end{align}
where $\Tilde{\Delta}_B = \mathtt{CSA}_{N \times N}^{-1} \Delta_B$. 

Keeping this equivalence in mind, the quantum CSA (QCSA) matrix can be defined as
\begin{align}
    &\mathtt{QCSA}_{N\times N}(f_{[L]},\alpha_{[N]},\beta_{[N]}) \nonumber\\ &\qquad=\diag(\beta_{[N]}) \mathtt{CSA}_{N\times N}(f_{[L]},\alpha_{[N]}).
\end{align}

To transition from the classical scheme to the quantum scheme, we need to restate a definition and \cite[Thm.~6]{nsumbox}.

\begin{definition}[Dual QCSA matrices]\label{dual_def}
    The matrices $H^u_N$ and $H^v_N$ are defined as $H^u_N =\mathtt{QCSA}_{N\times N}(f,\alpha,u)$ and $H^v_N =\mathtt{QCSA}_{N\times N}(f,\alpha,v)$. Then, $H^u_N$, and $H^v_N$ are dual QCSA matrices if the following are satisfied,
    \begin{enumerate}
        \item $u_1,\ldots,u_N$ are non-zero,
        \item $u_1,\ldots,u_N$ are distinct,
        \item for each $v_j,~j\in[1:N]$,
        \begin{align}
            v_j = \frac{1}{u_j} \left(\prod_{i=1 \atop i\neq j}^N(\alpha_j-\alpha_i)\right)^{-1}.
        \end{align}
    \end{enumerate}
\end{definition}

\begin{theorem}\label{feasile_nsum}
    For any dual QCSA matrices $H^u_N$, and $H^v_N$, there exists a feasible $N$-sum box transfer matrix $G(u,v)$ of size $N \times 2N$ given by,
    \begin{align}\label{tx-rx releation}
        G(u,v) = &\begin{bmatrix}
            I_L&0_{L\times\nu} & 0 & 0 & 0 & 0\\
            0 & 0 & I_{\mu-L} & 0 & 0 & 0\\ 0 & 0 & 0 & I_L & 0_{L\times\mu} & 0\\
            0 & 0 & 0 & 0 & 0 & I_{\nu-L}
            \end{bmatrix} \nonumber\\ &\times\begin{bmatrix}
            H^u_N & 0\\
            0& H^v_N
        \end{bmatrix}^{-1},
    \end{align}
    where $\nu = \lceil N/2\rceil$ and $\mu = \lfloor N/2\rfloor$.
\end{theorem}

Consider the first regime in Theorem~\ref{main_thm}, i.e., $X+T \geq \frac{N}{2}$. We note in \eqref{tx-rx releation} in Theorem~\ref{feasile_nsum} that there are $N$ symbols that are dropped. Thus, it is clear that the goal is to design a scheme such that we can drop interference terms from security and collusion while at the same time be able to decode the $L$ symbols. In the quantum setting the answers are generated for two instances, each with $L$ symbols, i.e., $S_n(1), S_n(2)$ store $2L$ symbols, each has $L$ symbols of all messages and they have the same structure as \eqref{classical_storage}, and $A_n^{[\theta]}(i)=S_n^t(i)Q_n^{[\theta]}$ for the honest servers. Thus, the answers can be written as $A^{[\theta]} (1)=H_N^uX(1)$, and $A^{[\theta]}(2)=H_N^vX(2)$, where $X(i)= [W_{\theta 1}(i),\ldots,W_{\theta L}(i), I'_1(i),\ldots,I'_{X+T}(i),0_{2B}]^t + \Tilde{\Delta}_B(i)$.
To see that decoding is possible, notice that after decoding-over-the-air is done, the received symbols are $G(u,v)\begin{bmatrix}
    A^{[\theta]}(1)&0\\
    0&A^{[\theta]}(2)
\end{bmatrix}$, the received answer terms for the $i$th instance, $i \in [2]$, will be of the form
\begin{align} \label{oneinstance}
   \begin{bmatrix}
     W_{\theta,1}+\sum_{k=1}^BC_{1,j(k)}U_k\\
        \vdots\\
        W_{\theta,L}+\sum_{k=1}^BC_{L,j(k)}U_k\\
        I'_{\gamma_i+1}+\sum_{k=1}^BC_{L+\gamma_i+1,j(k)}U_k\\
        \vdots \\
        I'_{X+T}+\sum_{k=1}^BC_{N-2B,j(k)}U_k\\ 
        \sum_{k=1}^BC_{N-2B+1,j(k)}U_k \\
        \vdots \\
        \sum_{k=1}^BC_{N,j(k)}U_k
 \end{bmatrix},
\end{align}
where $\gamma_1=X+T-\lfloor \frac{N}{2}\rfloor +L+2B=\lceil \frac{N}{2}\rceil$, $\gamma_2=X+T-\lceil \frac{N}{2}\rceil +L+2B=\lfloor \frac{N}{2}\rfloor$, $C$ is the inverse of the $\mathtt{CSA}_{N\times N}(f_{[L]},\alpha_{[N]})$, $U_i$ is the random noise variable due to the $i$th Byzantine server and $j:[B]\to [N]$ is a one-to-one mapping that is unknown to the user that maps the Byzantine server order to normal server order, i.e., $j(1)=5$ indicates that the 1st Byzantine server is the 5th server. Here, it should be noted that the Byzantine order does not really matter, i.e., if $Im(j_1)=Im(j_2)$, the effects are equivalent, where $Im(\cdot)$ defines the image of a function on its co-domain. 

Now, define
\begin{align}
    &R(a,b,j)=\begin{bmatrix}
        C_{1,j(1)} & \ldots & C_{1,j(b)} \\
        \vdots & \vdots & \vdots \\
        C_{a,j(1)} & \ldots & C_{a,j(b)}
    \end{bmatrix}, \\
        & T(j)=\begin{bmatrix}
        C_{L+X+T+1,j(1)} & \ldots & C_{L+X+T+1,j(B)} \\
        \vdots & \vdots & \vdots \\
        C_{L+X+T+B,j(1)} & \ldots & C_{L+X+T+B,j(B)}
    \end{bmatrix}, \\
        &Y(j)=\begin{bmatrix}
        C_{L+X+T+B+1,j(1)} & \ldots & C_{L+X+T+B+1,j(B)} \\
        \vdots & \vdots & \vdots \\
        C_{N,j(1)} & \ldots & C_{N,j(B)}
    \end{bmatrix}.
\end{align}

Note that \eqref{oneinstance} can be written in terms of $R(L,B,j)$, $T(j)$, and $Y(j)$ as follows
\begin{align}
    \begin{bmatrix}
        W_{\theta,1}\\
        \vdots\\
        W_{\theta,L}\\
        I'_{\gamma_i+1}+\sum_{k=1}^BC_{L+\gamma_i+1,j(k)}U_k\\
        \vdots \\
        I'_{X+T}+\sum_{k=1}^BC_{N-2B,j(k)}U_k\\ 
        0_{2B}
    \end{bmatrix}+\begin{bmatrix}
        R(L,B,j)\\
        0_{X+T-\gamma_i}\\
        T(j)\\
        Y(j)
    \end{bmatrix}\begin{bmatrix}
        U_1\\
        \vdots\\
        U_B
    \end{bmatrix}.
\end{align}

Next, we provide some important lemmas for the properties of these matrices that will help in the decoding stage. 

\begin{lemma}
\label{invertible}
    $Y(j)$ is always invertible. 
\end{lemma}

\begin{lemma} \label{decodelemma}
    Let $U=[U_1, \ldots, U_B]^t$. If 
    \begin{align} 
    T(j_1)Y^{-1}(j_1)Y(j_2)U=T(j_2)U,
    \end{align} 
    then, it follows that 
    \begin{align}
    \!\!\!\!R(N\!-\!2B, B, j_1)Y^{-1}(j_1)Y(j_2)U=R(N\!-\!2B, B, j_2)U.\!\!
    \end{align}
\end{lemma}

\begin{corollary} \label{finalcorollary}
    If 
    \begin{align}
    T(j_1)Y^{-1}(j_1)Y(j_2)U=T(j_2)U,
    \end{align} 
    then, it follows that 
    \begin{align} 
    R(L, B, j_1)Y^{-1}(j_1)Y(j_2)U=R(L, B, j_2)U.
    \end{align}
\end{corollary}

Then, notice that the last $B$ lines of \eqref{oneinstance} can be used to decode $U_i$ values, as they constitute $B$ linearly independent equations in $B$ variables, by using Lemma~\ref{invertible}. However, it should be noted again that $j$ is unknown to the user. Nevertheless, the user can guess a $j$. That is, let what the user guesses as the Byzantines $j_1$, and let $j_2$ be the real Byzantine servers. Then, let the user's estimate for the noise vector be  $\Tilde{U}$. Note that
\begin{align}
    \Tilde{U}=Y^{-1}(j_1)Y(j_2)U.
\end{align}
In the same manner, what is wanted is that the $\{L+X+T-\gamma_i+1,\ldots,N\}$th lines of \eqref{oneinstance} can be used for confirmation of whether the user predicted correct Byzantines, or $j_1=j_2$. That is, the user should check that
\begin{align}
    T(j_2)U=T(j_1)\Tilde{U}=T(j_1)Y^{-1}(j_1)Y(j_2)U.
\end{align}
If this is not true, it is guaranteed that $j_1 \neq j_2$, otherwise if it is satisfied, note that $R(L,B,j)U$ is exactly the noise effect on the first $L$ bits of \eqref{oneinstance} and Corollary~\ref{finalcorollary} shows that if the consistency check has been passed, then it does not matter whether the user has guessed correctly the Byzantine servers or not; for that noise value, the user guess is indistinguishable from the real Byzantines on the first $L$ bits of \eqref{oneinstance}. Thus, the user can just deduct the noise effects and get $\{W_{\theta,1},\ldots,W_{\theta,L}\}$.

For one instance, we showed using Lemma~\ref{decodelemma}, Corollary~\ref{finalcorollary}, and the discussion afterward that the consistency check does not require any of the interference terms in \eqref{oneinstance}. Thus, the presence of interference terms is insignificant in our decoding scheme, hence, we can drop them in this case. To that end, if $N$ is even and $X+T = \frac{N}{2}$, as an extreme example, \emph{all} the interference symbols can be dropped using the over-the-air decoding property and will not affect our decoding scheme.

In the second regime of Theorem \ref{main_thm}, the number of symbols that need to be dropped using the quantum scheme is greater than the interference terms, i.e., $\frac{N}{2}> X+T$. The main idea is to add extra interference, privacy, symbols such that $X+T' = \lceil \frac{N}{2} \rceil$ for the first retrieval instance, and $X+T'' = \lfloor \frac{N}{2} \rfloor$. Thus $L_1+L_2 = 2N - 2X-T'-T''-4B = N-4B$. If $N-4B > N-X-T-2B$, then it is better to use the quantum scheme, otherwise, using the classical scheme is better. In the third and final regime, after adding extra interference noise, there is not enough space for symbols, thus the classical scheme surpasses the quantum scheme in that regime.

After this point, the other ways for manipulating qudits by Byzantine servers, i.e., they can manipulate the encoding procedure, become much easier to deal with since they can be mapped to the sum of Pauli matrices. In Section~\ref{proofs}, we prove that our scheme is resilient against any qudit error. The proofs for security and privacy can be found in \cite{our_journal}.

\section{Generalization to Arbitrary Errors}\label{proofs} 
As discussed in the previous section, the proposed scheme can correct any Pauli error on a qudit. In this section, we show that this scheme can correct any error done on a qudit by the Byzantine server. Let us start with the following remark.

\begin{remark}
\label{byzantineq}
    Note that $\omega^{tr(ab)}\mathsf{X}(a)\mathsf{Z}(b)=\mathsf{Z}(b)\mathsf{X}(a)$ will accumulate a global phase in the quantum state it has been applied on, thus from a quantum state point of view, it can be discarded. Using this fact and one-time pad theorem, we note that there is an equivalence between a Byzantine server and an honest server whose link to the user is a noisy channel. We observe that the scheme that is resilient against the Byzantine server model is able to correct $B$ single qudit errors.
\end{remark}

\begin{lemma}
\label{basis}
    Let $M_{q}(\mathbb{C})$ be the vector space of all $q \times q$ matrices in field $\mathbb{C}$. Generalized Pauli matrices described in Section~\ref{problem_formulation} form a basis for this vector space.
\end{lemma}

\begin{Proof}
    Let $M=\sum_{a',b' \in \mathbb{F}_q}c_{a',b'}\ket{a'}\bra{b'}$ any arbitrary matrix in $M_q(\mathbb{C})$. It could equivalently be written as $M=\sum_{a,b \in \mathbb{F}_q}c_{a+b,b}\ket{a+b}\bra{b}$ using the transformation of variables $b=b',a=a'-b'$. Note that, this is indeed a valid transformation since it is invertible. The aim is to show that,
    \begin{align}
    \label{showthis}
        M&=\sum_{a,b \in \mathbb{F}_q}c_{a+b,b}\ket{a+b}\bra{b}=\sum_{k,l \in \mathbb{F}_q}u_{k,l}\mathsf{X}(k)\mathsf{Z}(l),
    \end{align}
    for unique $u_{k,l} \in \mathbb{C}$ values. To show this, note that
    \begin{align}
        \sum_{a,b \in \mathbb{F}_q}c_{a+b,b}&\ket{a+b}\bra{b}=\sum_{k,l\in \mathbb{F}_q}u_{k,l}\mathsf{X}(k)\mathsf{Z}(l)\\
        &=\sum_{k,l,m\in \mathbb{F}_q}u_{k,l}e^{i\frac{2\pi}{p}tr(ml)}\ket{m+k}\bra{m}.
    \end{align}
    Using the fact that $\ket{i}\bra{j}, i,j \in \mathbb{F}_q$ form a basis for $M_q(\mathbb{C})$, the equality condition transforms to,
    \begin{align}
    \label{four}
        c_{a+b,b}=\sum_{l\in\mathbb{F}_q}u_{a,l}e^{i\frac{2\pi}{p}tr(bl)}.
    \end{align}
    For fixed $a \in \mathbb{F}_q$, let $c=(c_{a+b,b})^t_{b\in\mathbb{F}_q}$ and $u=(u_{a,l})^t_{l \in \mathbb{F}_q}$ be the vectors. Then, from (\ref{four}), we have,   $c=Fu$, where $F_{kl}=e^{i\frac{2\pi}{p}tr(kl)}$. From \cite{hayashibook}, we note that this matrix describes the Fourier basis coefficients in $\mathbb{F}_q$, thus is an invertible matrix. Therefore, from $c$, a unique $u$ can be found for each $a \in \mathbb{F}_q$. Thus, (\ref{showthis}) is satisfied indeed with unique $u_{k,l}$ values.
\end{Proof}

\begin{corollary}\label{krauscorollary}
    Using Kraus operator form of the quantum operations of Definition~\ref{kraus}, any quantum operation $\mathcal{E}$ with operation elements $\{E_i\}_i$ can be written as a linear combination of generalized Pauli operators.
\end{corollary}

\begin{remark}
    It is important to note that the Byzantine servers can only send one qudit to the user, i.e., the Hilbert space dimensions do not change after applying the quantum operations.
\end{remark}

\begin{lemma}[Theorem~10.2 in \cite{nielsen-chuang}]\label{lincom}
    Let $\mathcal{E}$ be a quantum operation with elements $\{E_i\}_i$ and let $C$ be a quantum code that satisfies the quantum error correction condition. Then, $C$ also satisfies the quantum error correction condition for the quantum operation $\mathcal{F}$ with operation elements $\{F_i\}_i$ where $F_i=\sum_{i,j}m_{ij}E_i$, where $m_{ij} \in \mathbb{C}$. This is also known as ``discretization of quantum errors.''
\end{lemma}

Thus, it follows from Remark~\ref{byzantineq}, Lemma~\ref{basis}, and Corollary~\ref{krauscorollary} that any operation done by the individual Byzantine servers on their individual qudits can be written in terms of Kraus representation. By using Lemma~\ref{lincom}, since our scheme protects against Pauli errors, it can protect against any errors by the Byzantine servers of their qudits.

\newpage

\bibliographystyle{unsrt}
\bibliography{references.bib}

\begin{thebibliography}{10}

\bibitem{chor}
B.~Chor, E.~Kushilevitz, O.~Goldreich, and M.~Sudan.
\newblock Private information retrieval.
\newblock {\em Jour. of the ACM}, 45(6):965--981, November 1998.

\bibitem{c_pir}
H.~Sun and S.~A. Jafar.
\newblock The capacity of private information retrieval.
\newblock {\em IEEE Trans. Info. Theory}, 63(7):4075--4088, July 2017.

\bibitem{colluding}
H.~Sun and S.~A. Jafar.
\newblock The capacity of robust private information retrieval with colluding databases.
\newblock {\em IEEE Trans. Info. Theory}, 64(4):2361--2370, April 2018.

\bibitem{first_xsecure}
H.~Yang, W.~Shin, and J.~Lee.
\newblock Private information retrieval for secure distributed storage systems.
\newblock {\em IEEE Trans. Info. Foren. Security}, 13(12):2953--2964, May 2018.

\bibitem{csa}
Z.~Jia, H.~Sun, and S.~A. Jafar.
\newblock Cross subspace alignment and the asymptotic capacity of {$X$}-secure {$T$}-private information retrieval.
\newblock {\em IEEE Trans. Info. Theory}, 65(9):5783--5798, May 2019.

\bibitem{byzantine_tpir}
K.~Banawan and S.~Ulukus.
\newblock The capacity of private information retrieval from {B}yzantine and colluding databases.
\newblock {\em IEEE Trans. Info. Theory}, 65(2):1206--1219, September 2018.

\bibitem{jafar_byzantine}
Z.~Jia and S.~Jafar.
\newblock X-secure t-private information retrieval from {MDS} coded storage with {B}yzantine and unresponsive servers.
\newblock {\em IEEE Transactions on Information Theory}, 66(12):7427--7438, July 2020.

\bibitem{arbitrarycollusion}
X.~Yao, N.~Liu, and W.~Kang.
\newblock The capacity of private information retrieval under arbitrary collusion patterns for replicated databases.
\newblock {\em IEEE Trans. Info. Theory}, 67(10):6841--6855, July 2021.

\bibitem{banawan_eaves}
K.~Banawan and S.~Ulukus.
\newblock Private information retrieval through wiretap channel {II}: Privacy meets security.
\newblock {\em IEEE Trans. Info. Theory}, 66(7):4129--4149, February 2020.

\bibitem{banawan_multimessage_pir}
K.~Banawan and S.~Ulukus.
\newblock Multi-message private information retrieval: Capacity results and near-optimal schemes.
\newblock {\em IEEE Trans. Info. Theory}, 64(10):6842--6862, April 2018.

\bibitem{banawan_pir_mdscoded}
K.~Banawan and S.~Ulukus.
\newblock The capacity of private information retrieval from coded databases.
\newblock {\em IEEE Trans. Info. Theory}, 64(3):1945--1956, January 2018.

\bibitem{batuhan_hetero}
K.~Banawan, B.~Arasli, Y.-P. Wei, and S.~Ulukus.
\newblock The capacity of private information retrieval from heterogeneous uncoded caching databases.
\newblock {\em IEEE Trans. Info. Theory}, 66(6):3407--3416, June 2020.

\bibitem{C_SETPIR}
Q.~Wang, H.~Sun, and M.~Skoglund.
\newblock The capacity of private information retrieval with eavesdroppers.
\newblock {\em IEEE Transactions on Information Theory}, 65(5):3198--3214, December 2018.

\bibitem{ChaoTian}
C.~Tian, H.~Sun, and J.~Chen.
\newblock Capacity-achieving private information retrieval codes with optimal message size and upload cost.
\newblock {\em IEEE Trans. Info. Theory}, 65(11):7613--7627, November 2019.

\bibitem{codedstorage_adversary_tpir}
L.~Holzbaurand, R.~Freij-Hollanti, and C.~Hollanti.
\newblock On the capacity of private information retrieval from coded, colluding, and adversarial servers.
\newblock In {\em IEEE ITW}, August 2019.

\bibitem{grpahbased_pir}
N.~Raviv, I.~Tamo, and E.~Yaakobi.
\newblock Private information retrieval in graph-based replication systems.
\newblock {\em IEEE Trans. Info. Theory}, 66(6):3590--3602, November 2019.

\bibitem{Kumar_PIRarbCoded}
S.~Kumar, H.-Y. Lin, E.~Rosnes, and A.~G. i~Amat.
\newblock Achieving maximum distance separable private information retrieval capacity with linear codes.
\newblock {\em IEEE Trans. Info. Theory}, 65(7):4243--4273, July 2019.

\bibitem{multimessage_pir_sideinfo}
M.~J. Siavoshani, S.~P. Shariatpanahi, and M.~Ali Maddah-Ali.
\newblock Private information retrieval for a multi-message scenario with private side information.
\newblock {\em IEEE Trans. Commun.}, 69(5):3235--3244, January 2021.

\bibitem{mat_pir_byzantine_colluding}
P.~Saarela, M.~Allaix, R.~Freij-Hollanti, and C.~Hollanti.
\newblock Private information retrieval from colluding and {B}yzantine servers with binary {R}eed–{M}uller codes.
\newblock In {\em IEEE ISIT}, June 2022.

\bibitem{mdstpir}
H.~Sun and S.~A. Jafar.
\newblock Private information retrieval from {MDS} coded data with colluding servers: Settling a conjecture by {F}reij-{H}ollanti et al.
\newblock {\em IEEE Trans. Info. Theory}, 64(2):1000--1022, December 2017.

\bibitem{multiround_pir}
H.~Sun and S.~A. Jafar.
\newblock Multiround private information retrieval: Capacity and storage overhead.
\newblock {\em IEEE Trans. Info. Theory}, 64(8):5743--5754, January 2018.

\bibitem{nomeirasymmetric}
M.~Nomeir, S.~Vithana, and S.~Ulukus.
\newblock Asymmetric {$X$}-secure {$T$}-private information retrieval: More databases is not always better.
\newblock Available online at arXiv:2305.05649.

\bibitem{UlukusPIRLC}
S.~Ulukus, S.~Avestimehr, M.~Gastpar, S.~A. Jafar, R.~Tandon, and C.~Tian.
\newblock Private retrieval, computing, and learning: Recent progress and future challenges.
\newblock {\em IEEE Journal on Selected Areas in Communications}, 40(3):729--748, March 2022.

\bibitem{qpir}
S.~Song and M.~Hayashi.
\newblock Capacity of quantum private information retrieval with multiple servers.
\newblock {\em IEEE Trans. Info. Theory}, 67(1):452--463, September 2021.

\bibitem{qtpir}
S.~Song and M.~Hayashi.
\newblock Capacity of quantum private information retrieval with colluding servers.
\newblock {\em IEEE Transactions on Information Theory}, 67(8):5491--5508, May 2021.

\bibitem{nsumbox}
M.~Allaix, Y.~Lu, Y.~Yao, T.~Pllaha, C.~Hollanti, and S.~A. Jafar.
\newblock {$N$}-sum box: An abstraction for linear computation over many-to-one quantum networks.
\newblock 2023.
\newblock Available online at arXiv: 2304.07561.

\bibitem{qpir_colluding_mdscoded}
M.~Allaix, S.~Song, L.~Holzbaur, T.~Pllaha, M.~Hayashi, and C.~Hollanti.
\newblock On the capacity of quantum private information retrieval from {MDS}-coded and colluding servers.
\newblock {\em IEEE Jour. Sel. Areas Commun.}, 40(3):885--898, January 2022.

\bibitem{qpir_star_product_codes}
M.~Allaix, L.~Holzbaur, T.~Pllaha, and C.~Hollanti.
\newblock High-rate quantum private information retrieval with weakly self-dual star product codes.
\newblock In {\em IEEE ISIT}, July 2021.

\bibitem{qtpir_t=n-1}
S.~Song and M.~Hayashi.
\newblock Capacity of quantum symmetric private information retrieval with collusion of all but one of servers.
\newblock {\em IEEE Jour. Sel. Areas Info. Theory}, 2(1):380--390, January 2021.

\bibitem{our_journal}
A.~Aytekin, M.~Nomeir, S.~Vithana, and S.~Ulukus.
\newblock {Quantum $X$-Secure $E$-Eavesdropped $T$-Colluding Symmetric Private Information Retrieval}.
\newblock Available online at https://user.eng.umd.edu/~ulukus/papers/journal/QXSETSPIR.pdf.

\bibitem{nielsen-chuang}
M.~Nielsen and I.~Chuang.
\newblock {\em Quantum Computation and Quantum Information: 10th Anniversary Edition}.
\newblock Cambridge University Press, 2010.

\bibitem{our_quantum_first}
A.~Aytekin, M.~Nomeir, S.~Vithana, and S.~Ulukus.
\newblock Quantum symmetric private information retrieval with secure storage and eavesdroppers.
\newblock In {\em IEEE Globecom}, December 2023.

\bibitem{hayashibook}
M.~Hayashi.
\newblock {\em Group Representation for Quantum Theory}.
\newblock Springer, 2017.

\end{thebibliography}

\end{document}